\documentclass[aps,prl,longbibliography,reprint,showpacs,floatfix,superscriptaddress]{revtex4-1}
\usepackage{graphicx}
\usepackage{amsmath,amssymb,amsfonts}
\usepackage{hyperref}

\begin{document}

\title{Nonvolatile Solid-State Charged-Polymer Gating of Topological Insulators into the Topological Insulating Regime}

\author{R. M. Ireland}
\affiliation{Department of Materials Science and Engineering and Department of Chemistry, The Johns Hopkins University, Baltimore, MD, USA 21218}

\author{Liang Wu}
\affiliation{Department of Physics and Astronomy, The Johns Hopkins University, Baltimore, MD 21218, USA}

\author{M. Salehi}
\affiliation{Department of Material Science and Engineering, Rutgers, The State University of New Jersey, Piscataway, NJ, 08854, USA }

\author{S. Oh}
\affiliation{Department of Physics and Astronomy, Rutgers, The State University of New Jersey, Piscataway, NJ, 08854, USA}

\author{N. P. Armitage}
\affiliation{Department of Physics and Astronomy, The Johns Hopkins University, Baltimore, MD 21218, USA}

\author{H. E. Katz}
\affiliation{Department of Materials Science and Engineering and Department of Chemistry, The Johns Hopkins University, Baltimore, MD, USA 21218}

\date{\today}

\begin{abstract}

We demonstrate the ability to reduce the carrier concentration of thin films of the topological insulator (TI) Bi$_2$Se$_3$ by utilizing a novel approach, namely non-volatile electrostatic gating via corona charging of electret polymers. Sufficient electric field can be imparted to a polymer-TI bilayer to result in significant electron density depletion, even without the continuous connection of a gate electrode or the chemical modification of the TI.  We show that the Fermi level of Bi$_2$Se$_3$ is shifted towards the Dirac point with this method. Using THz spectroscopy, we find that the surface chemical potential is lowered into the bulk band gap ($\sim$ 50 meV above the Dirac point  and 170 meV below the conduction band minimum) and it is stabilized in the intrinsic regime while enhancing electron mobility. This represents the first use of a charged polymer gate for modulating TI charge density.  The mobility of surface state electrons is enhanced to a value as high as $\sim$1600 cm$^2$/Vs at 5K. 

\end{abstract}

\maketitle

\section{Introduction}

Topological insulators (TIs) are a recently discovered class of materials, which support helical metallic surface states despite being (ideally) bulk insulators \cite{Hasan-Moore-10,HasanKaneRMP10,Qi-Zhang-11}. Unlike conventional surface states, their existence does not depend of the details of surface termination or environment, but only on certain properties of the underlying quantum mechanical wavefunctions.  Examples of such materials are Bi$_{1-x}$Sb$_x$,  Bi$_2$Se$_3$, Bi$_2$Te$_3$, and TlBiSe$_2$. The spin direction in the surface states is constrained to be in the plane of the surface and perpendicular to the momentum and therefore they exhibit suppressed backscattering when carrying a current. This novel spin structure may make TIs a novel platform for future low-dissipation spintronics applications \cite{Pesin12a}. Topological surface states may also display a host of interesting fundamental physics phenomena including a topological magneto-electric effect, quantized Faraday rotation and axion electrodynamics \cite{maciejko2010fractional,qi2009inducing}. Early on, photoemission confirmed \cite{HsiehNature08,ChenScience09,HsiehNature09} the existence of the surface states, however further development of the field hinges on controlling bulk electronic properties sufficiently to allow the surface properties to become dominant.

In this regard, the unique properties of TIs can only be fully realized in intrinsic (i.e. bulk-insulating) TIs.  Reaching the intrinsic regime depends crucially on positioning the electrochemical potential in the bulk band gap.  However an outstanding issue in this field is that most as-grown materials like  Bi$_2$Se$_3$ typically have their chemical potential in the conduction band due to defect doping or band bending effects at surfaces, so their bulk properties are decidedly not insulating \cite{AnalytisPRB2010}.  This includes the high quality molecular beam epitaxy (MBE)-grown Bi$_2$Se$_3$ used in much of our own earlier work\cite{ValdesAguilarPRL12, BansalPRL12, ValdesAguilarJAP13, WuNatPhys13}. Although MBE is the most precise method \cite{brahlek2011surface} for fabricating high quality  Bi$_2$Se$_3$, most films prepared by this method will still have a high intrinsic n-type carrier concentration due to Se vacancies.  In order to utilize the intrinsic TI physics, it is necessary to further deplete negative carriers.

Although there have been a number of recent advances \cite{Koirala15a,Wu16b} in growing TIs with the chemical potential in the gap, further progress is needed to be able to tune the Fermi energy.  Chemical potential manipulation may be achieved by means of a capacitively coupled gate electrode.  For example, one may couple the gate electrode to the face of a two-dimensional inorganic sample via an ionic liquid \cite{Segawa12a,Shimizu12a,Yuan11a}. This configuration, while indeed conveying a high capacitive field (up to 5 V across a molecular distance, $\sim$10 MV/cm) to the sample, suffers from a number of problems, including chemical inhomogeneity at the inorganic surface (a source of hysteresis and mobility-reducing scattering), limited control of parasitic capacitances and resistances, the possibility of electrochemistry at the inorganic sample surface, and an inability to confine the entire device structure to dimensions that would allow interference-free optical probing.

Some of these issues are ameliorated using solid gates.  For example, ceramic \cite{Chang15a,Chen10a,Kim12a} gates have been used to observe changes in conductivity.  However, this still requires connection and operation of a gate electrode to change the density during transport measurements.  To avoid this, a few quasipermanent gating modifications have been made to TIs and related materials.   For instance, it was shown that electron irradiation reversed the majority carrier sign in bismuth telluride \cite{Rischau13a}.  In a related demonstration to the current work, corona charging of graphene on silicon carbide tuned the charge density through the Dirac point \cite{Lartsev14a}.  Electron-accepting overlayers such as molybdenum trioxide and tetracyanoquinodimethane can be deposited adjacent to TIs \cite{Wu16a}.  A recent work also utilized organics to engineer topological insulators, but they increase the Fermi energy and created Rashba surface states, which are topologically trivial \cite{Jakobs15a}.

In this work, we demonstrate the ability to tailor the carrier concentration of  Bi$_2$Se$_3$ films prepared by MBE by utilizing an approach without precedent in the TI field: nonvolatile electrostatic gating via the corona charging of electret polymers \cite{Giacometti99a}.  Sufficient electric field can be imparted to a polymer-TI bilayer to result in significant electron depletion without the continuous connection of a gate electrode or the chemical modification of the TI.  We show that the Fermi level of  Bi$_2$Se$_3$ is shifted towards the Dirac point by altering  Bi$_2$Se$_3$ carrier concentrations with this method. We find that electrostatic gating is highly effective for lowering the surface chemical potential into the bulk band gap and stabilizing it in the intrinsic regime while enhancing electron mobility.

\begin{figure}[tb]
\includegraphics[width=0.8\columnwidth]{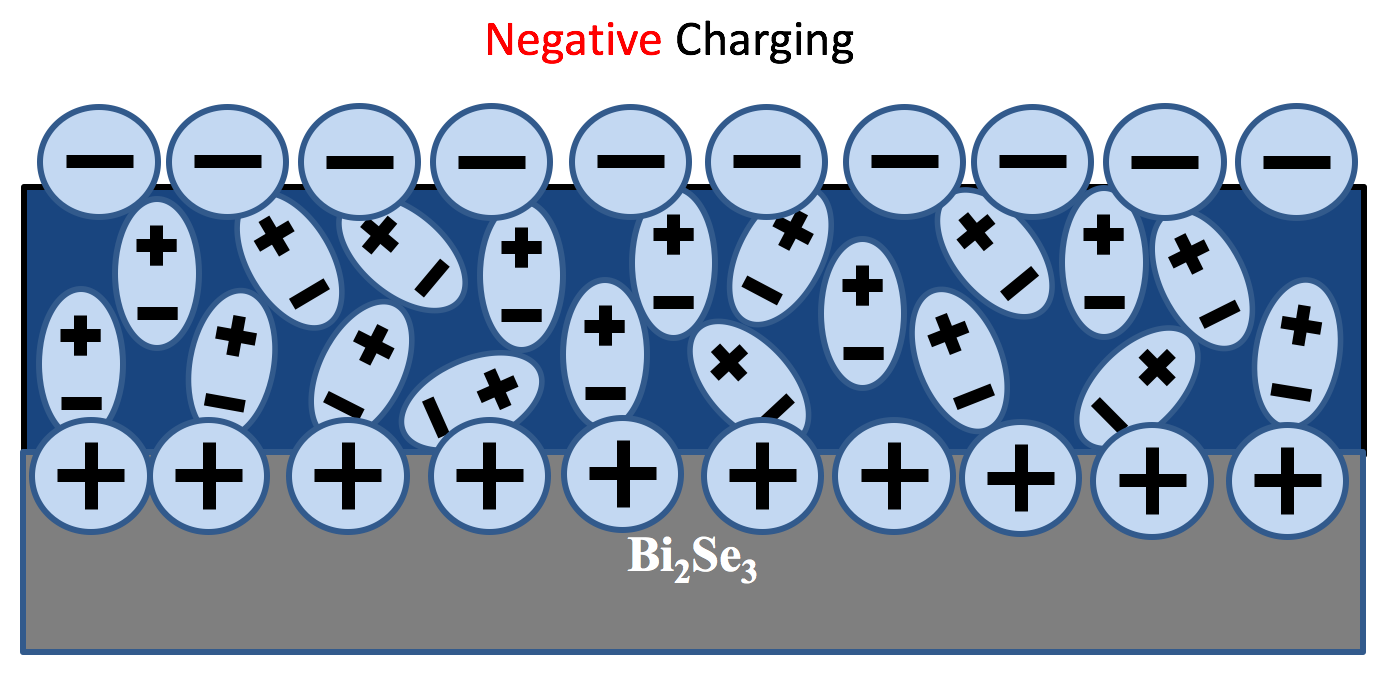}
\caption{Schematic of charging of polymer film on Bi$_2$Se$_3$ with negative ions that results in electron depletion of TI films.  Blue represents the charged polymer layer.  Polarized dipoles in the bulk of the polymer are also indicated.  }
\label{ChargingCartoon}
\end{figure}

A corona discharge is an electrical discharge that occurs when a high voltage is applied between asymmetric electrodes such as a point and a plate.  The emission or extraction of electrons induced by high voltage at the point creates gas-phase ions that move through a region in space, are brought to a defined potential by passing through a metal grid electrode set to that potential, and eventually become adsorbed onto or embedded near the surface of a polymer placed near the plate, a process that can be termed field-assisted adsorption. In contrast to techniques like electron-beam poling, in which electrons strike a surface with keV energies and penetrate into the bulk of the polymer, the energy of the corona ions is much smaller as they first move through the ambient atmosphere and then become equilibrated to the grid.    The ions do not penetrate appreciably into the bulk of the materials, but instead transfer their charge to the surface where it resides or is injected into bulk traps as shown in Fig. \ref{ChargingCartoon} for negative charging, which results in depletion of a TI film.  It has been used for applications in electrophotography, charging of electrostatic filters, and the poling of polymers to induce ferroelectricity and nonlinear optical effects \cite{Giacometti99a}.   In the present case we use it to charge a polymer electret and build-in a stable non-volatile electric field that can reduce the charge density of the TI surface states that are in contact with the polymer.

We selected commercially available or easily synthesized polymers: polystyrene (PS), polystyrene with 3$\%$ chloromethylstyrene comonomer (PS-Cl), polystyrene with 3$\%$ C60-methyl side chains (PS-C60)\cite{alley2016synthesis} (made by direct substitution of the chloromethyl group with C60), and the perfluorinated polymer Novec. All four are highly hydrophobic, have wide electrochemical windows, and would be expected to have little or no chemical interaction with  Bi$_2$Se$_3$.   Polystyrene is a known electret/charge storage material, with demonstrated capability to tune semiconductor charge densities in a nonvolatile fashion. \cite{Dawidczyk14a,Dawidczyk12a,Erhard10a,Reuter10a}.  The chloromethylstyrene and C60-methylstyrene derivatives were investigated to see whether the labile chloro groups or additional fullerene energy levels would increase charge trapping relative to polystyrene.  Novec is a non-ionic polymeric fluorochemical surfactant in a proprietary nonafluorobutyl ether that has not been used for the present purpose before, but as shown below, is particularly advantageous for the application under consideration.

\begin{figure}[tb]
\includegraphics[width=0.75\columnwidth]{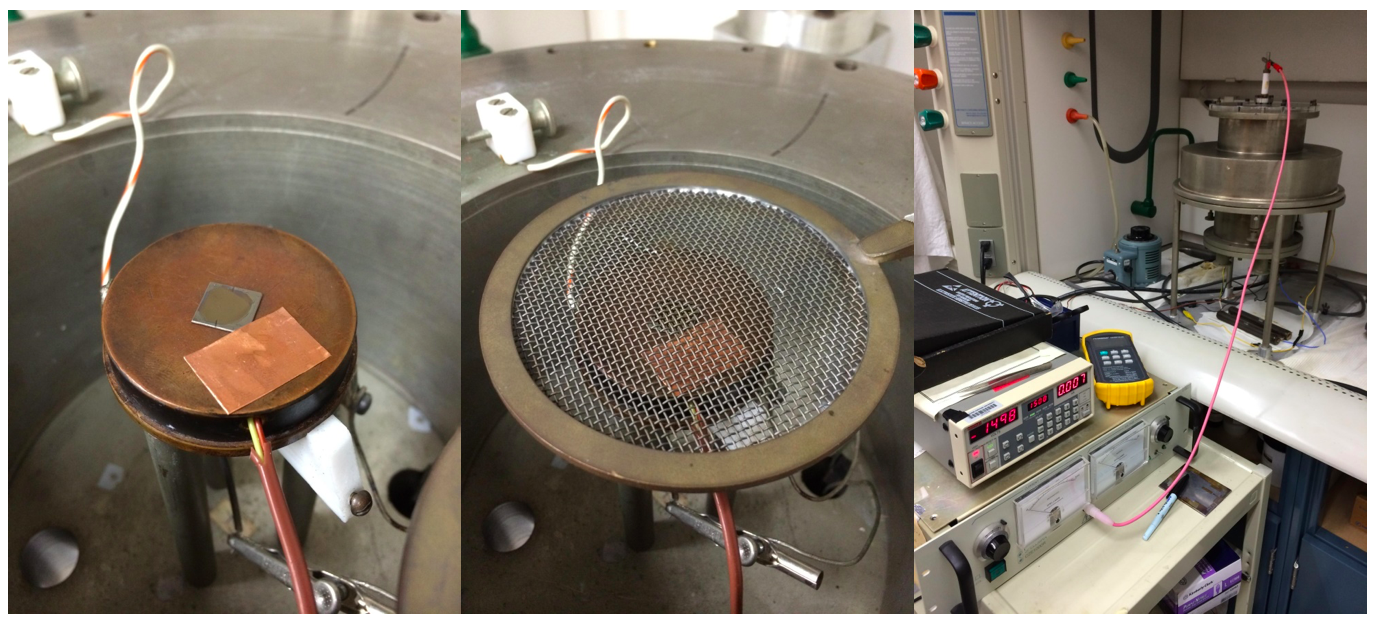}
\includegraphics[width=0.75\columnwidth]{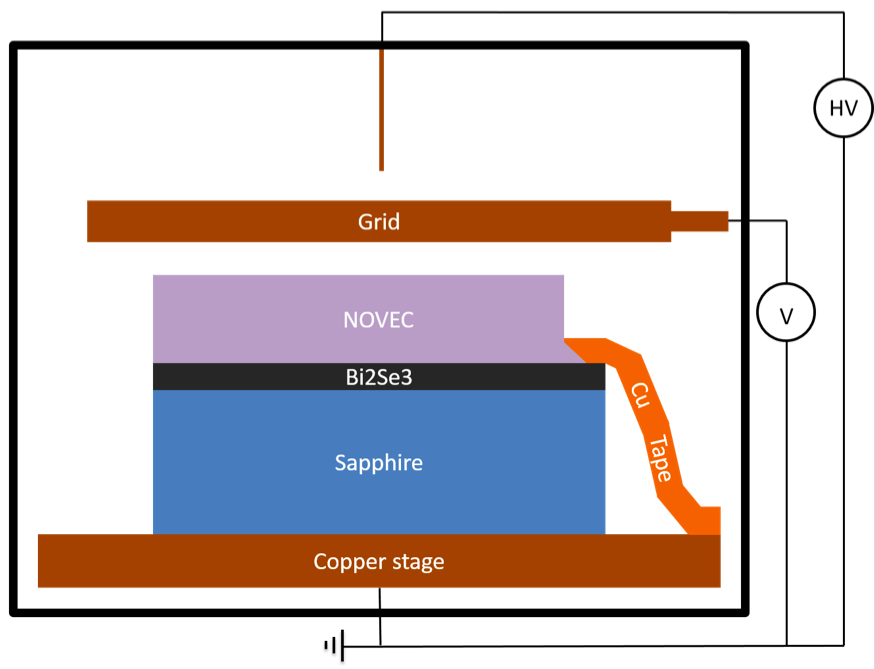}
\caption{Images (top, left to right) of a sample and copper tape on the copper stage, with the grid screen in place, and the external connection to the needle corona source. Schematics (bottom) of the corona charging setup with copper tape grounding of the  Bi$_2$Se$_3$ layer. }
\label{Fig1}
\end{figure}

 \section{Experimental Details}

Thin films in this study were 16 QL thick (1 QL is a unit cell and corresponds to $\approx$ 1 nm) low-carrier-density Bi$_2$Se$_3$ molecular beam epitaxy films grown on Bi$_{2-x}$In$_x$Se$_3$ buffer layers grown on sapphire substrates using the Rutgers University custom-designed molecular beam epitaxy (MBE) system.  Details are published elsewhere \cite{Koirala15a}.  DC measurements, photoemission spectroscopy, and THz experiments have shown no signatures of bulk or trivial two dimensional electron gas states \cite{Koirala15a}.  After growth, samples were sealed in vacuum bags and sent to Johns Hopkins University overnight.

\begin{figure}[tb]
\includegraphics[width=0.8\columnwidth,angle=0]{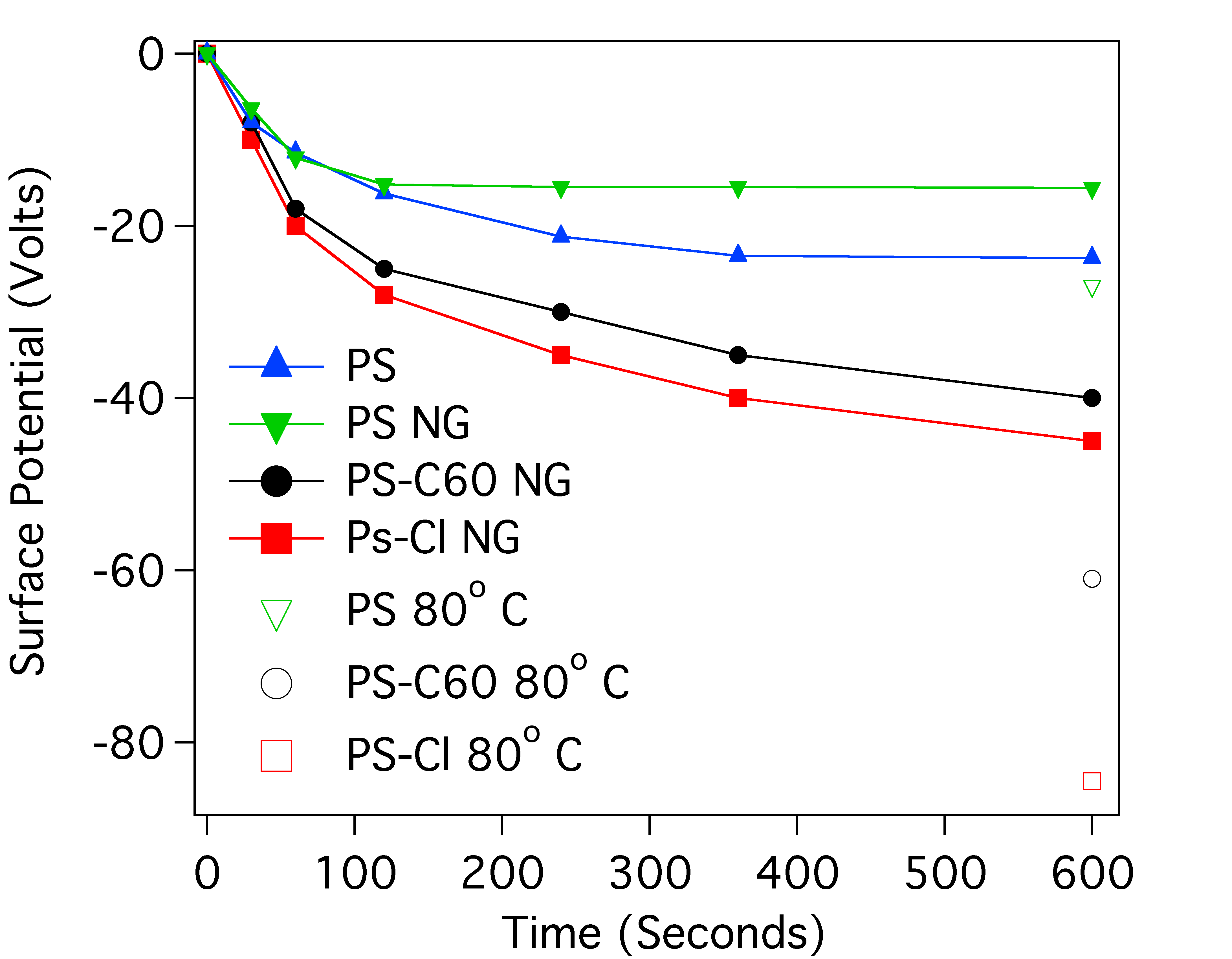}
\includegraphics[width=0.8\columnwidth,angle=0]{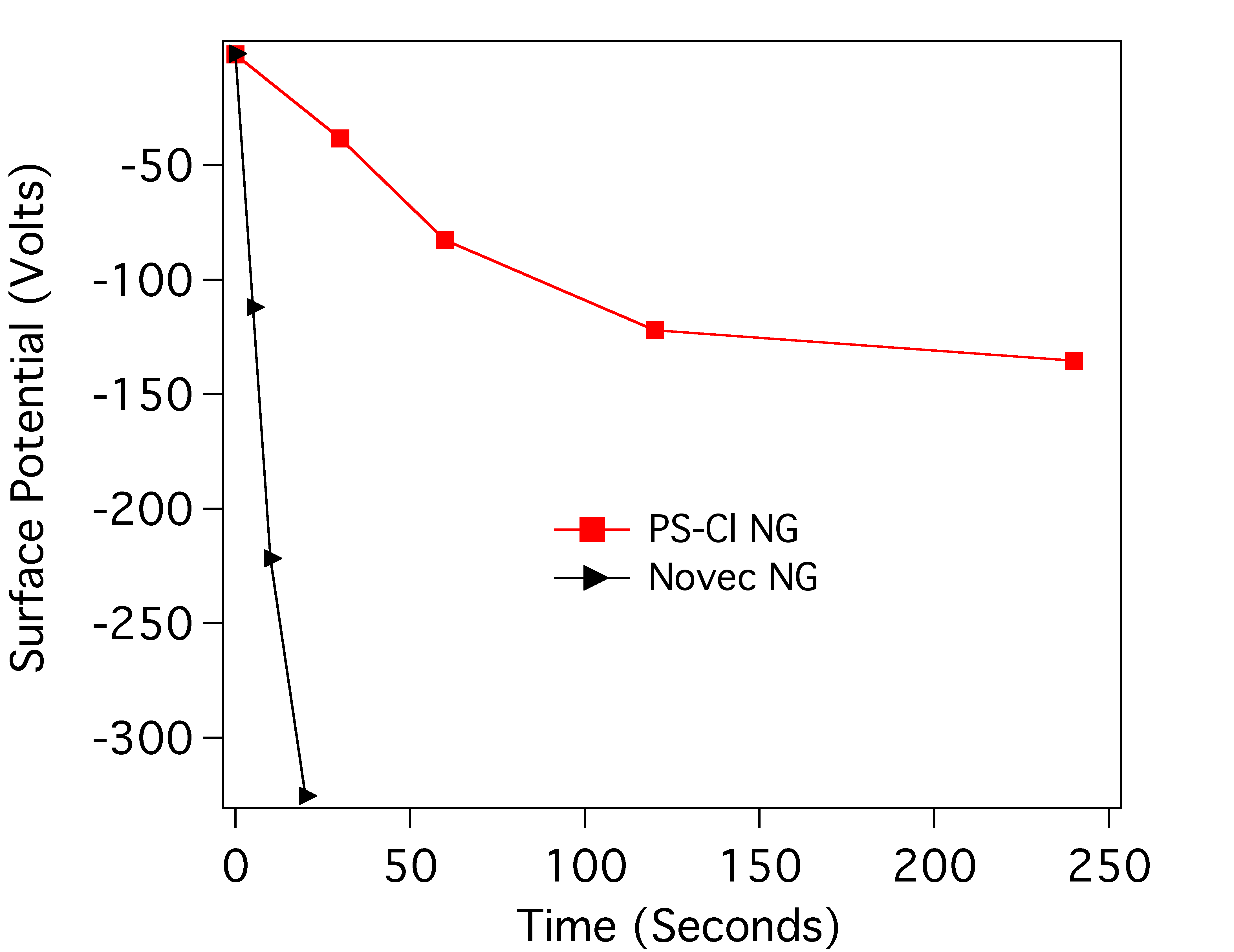}
\caption{(top) Surface voltage as a result of corona charging as a function of time with -1500 V applied to the grid over spin-coated 0.1 micrometer thick polymer films deposited on Au films at.  Shown is data for polystyrene (PS), polystyrene with 3$\%$ chloromethylstyrene comonomer (PS-Cl), polystyrene with 3$\%$ C60-methyl side chains (PS-C60), and the perfluorinated polymer Novec.  NG (not grounded) refers to samples that were not connected between the gold and the grounded copper stage.  80$^\circ$ C refers to final surface voltages after heating the samples to 80$^\circ$ C during the charging.  (bottom) Surface voltage measured for a drop-coated polymer  (thickness  $\sim$ 5 $ \mu$m) deposited on Bi$_2$Se$_3$ films at -1500 V grid voltage.}
\label{SurfaceCharge}
\end{figure}

Some polymers were drop-cast directly from solution onto
substrates with premade Bi$_2$Se$_3$ films using about 0.1 ml
of solution per square centimeter of substrate area. The
polymer solvent is allowed to evaporate, resulting in polymer
film thicknesses around 2-5 micrometers.  Alternatively the 
films are spin-cast at 1000 rpm for 60 seconds in order
to reduce the thickness and drying times.   A single spin-cast layer is typically about 0.1 micrometers. Novec dries
in a matter of minutes, whereas the polystyrene-based
polymers are cast from chlorobenzene solutions diluted
to 5 mg/ml, so drop-cast films may take more than
12 hours to dry in ambient atmosphere.   As it dries quickly spin-cast Novec layers can be built up in repeated depositions, whereas polystyrene is too soluble to efficiently build up in this fashion.  All  films
are further dried under low vacuum at 60$^\circ$ C for a couple of
hours.  The recommended environment for applying Novec coating is 20-27$^\circ$ C and 40-70 $\%$ relative humidity to prevent moisture condensation on the polymer film and provide a reasonable drying time.

 \section{Charging Studies}

We performed a systematic study of conditions and materials that would allow the maximum static electric field to be stored in a polymer dielectric layer overlaying the Bi$_2$Se$_3$.  Static charge was applied to the upper polymer surface via a metal screen (the ``grid") powered by a corona (plasma) emanating from a needle above the grid, establishing a potential of 10,000 volts between the needle and a grounded copper stage, as shown in Fig. \ref{Fig1}.  For some experiments, copper tape was used to connect the grounded stage to the  Bi$_2$Se$_3$ in an attempt to set the  Bi$_2$Se$_3$ semiconducting layers to ground as well.  A triode arrangement with a grid is used to ensure greater uniformity of potential over the sample as previously demonstrated \cite{Giacometti99a}.  A grounded metal chamber encloses the entire apparatus, and is not removed when the needle or grid are powered.   The stored voltage depended both on the charging time and grid voltage.  The stored voltage was indicated by measuring with an electrostatic voltmeter. The electrostatic voltmeter is a surface DC voltmeter that measures surface potential in the range of $\pm$3000 V without physically contacting the measured surface. It features high-accuracy (0.1$\%$), drift-free measurements that are almost fully independent of probe-to-surface separation.

 \begin{figure*}[tb]
\includegraphics[width=0.65\columnwidth]{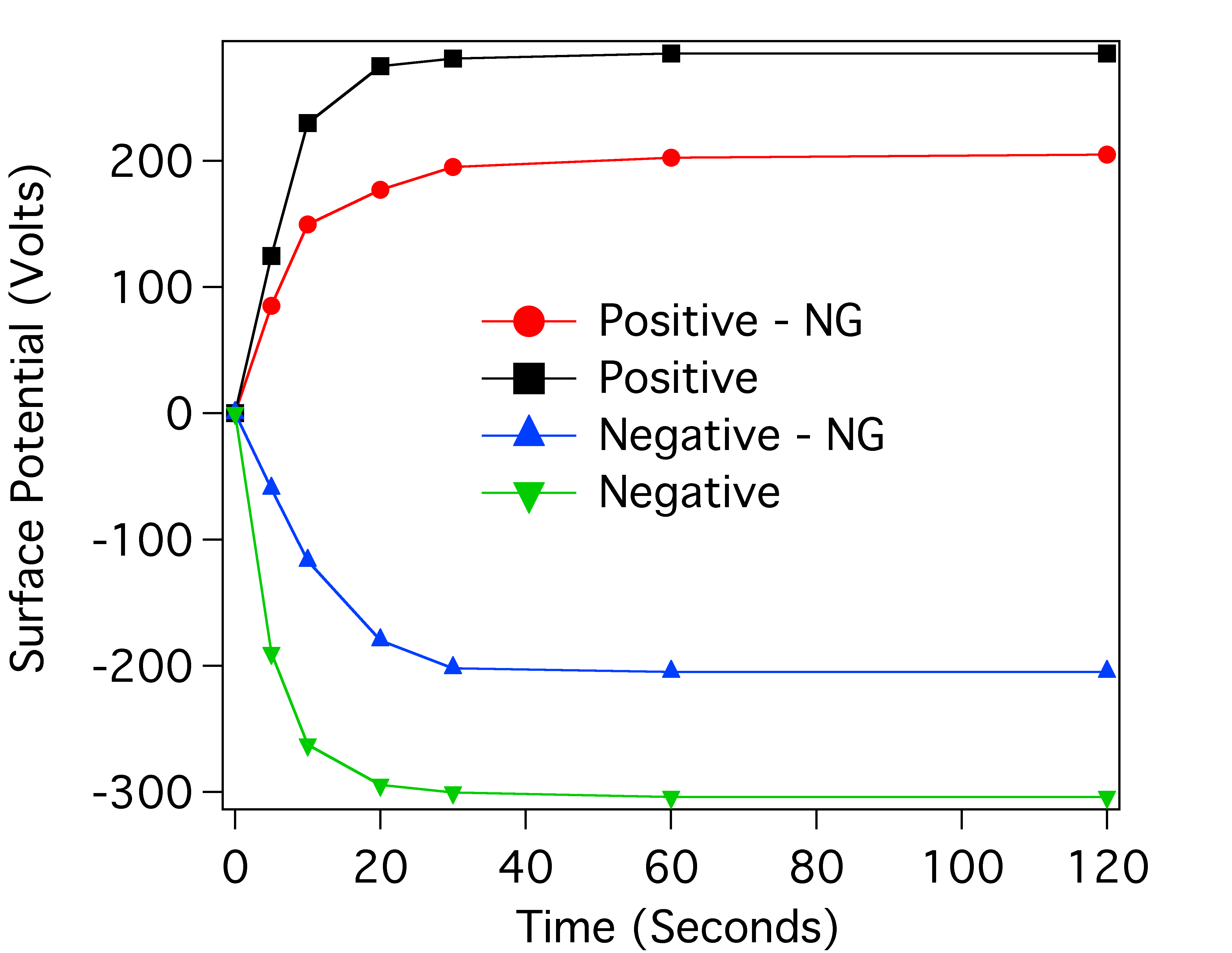}
\includegraphics[width=0.65\columnwidth]{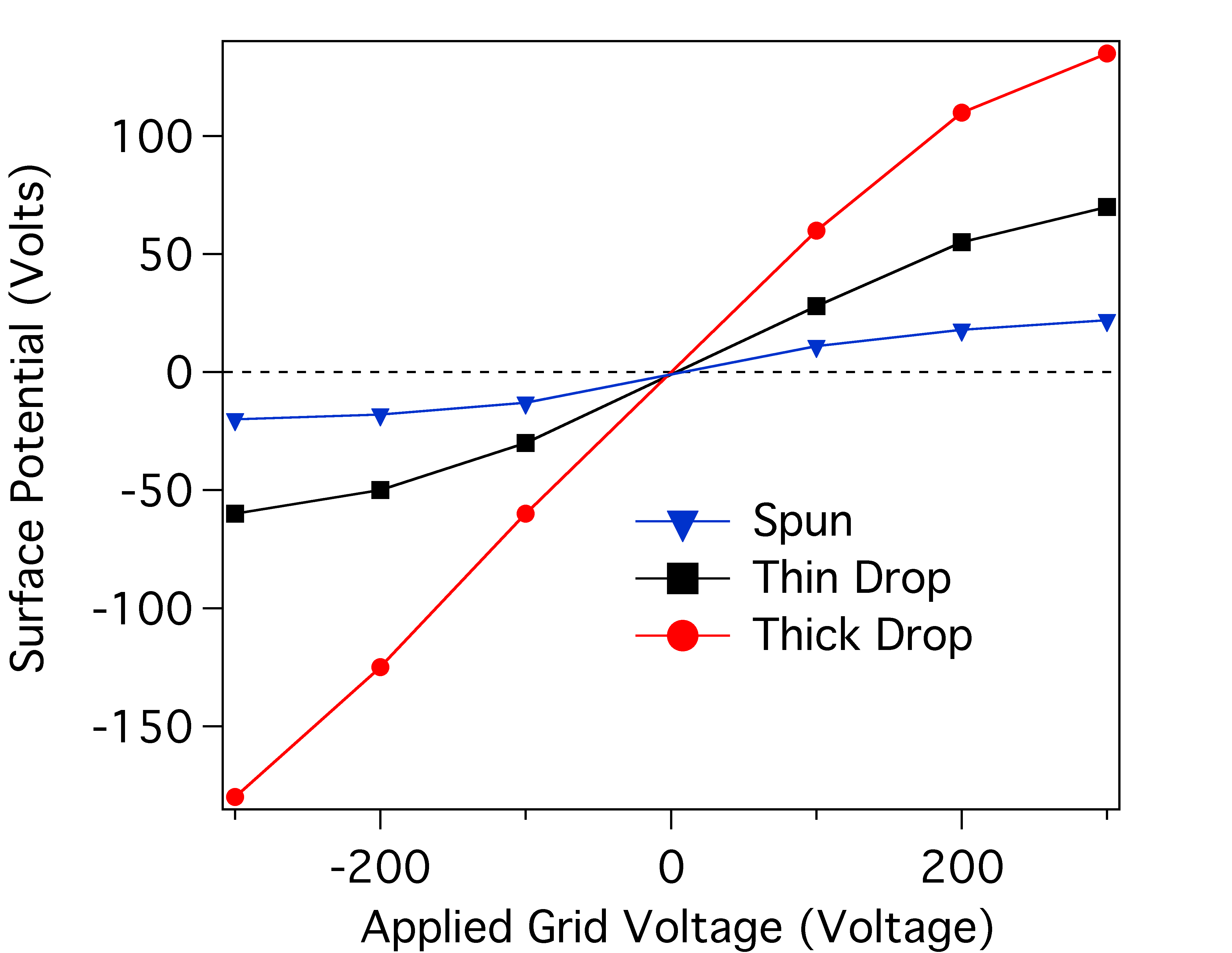}
\includegraphics[width=0.65\columnwidth]{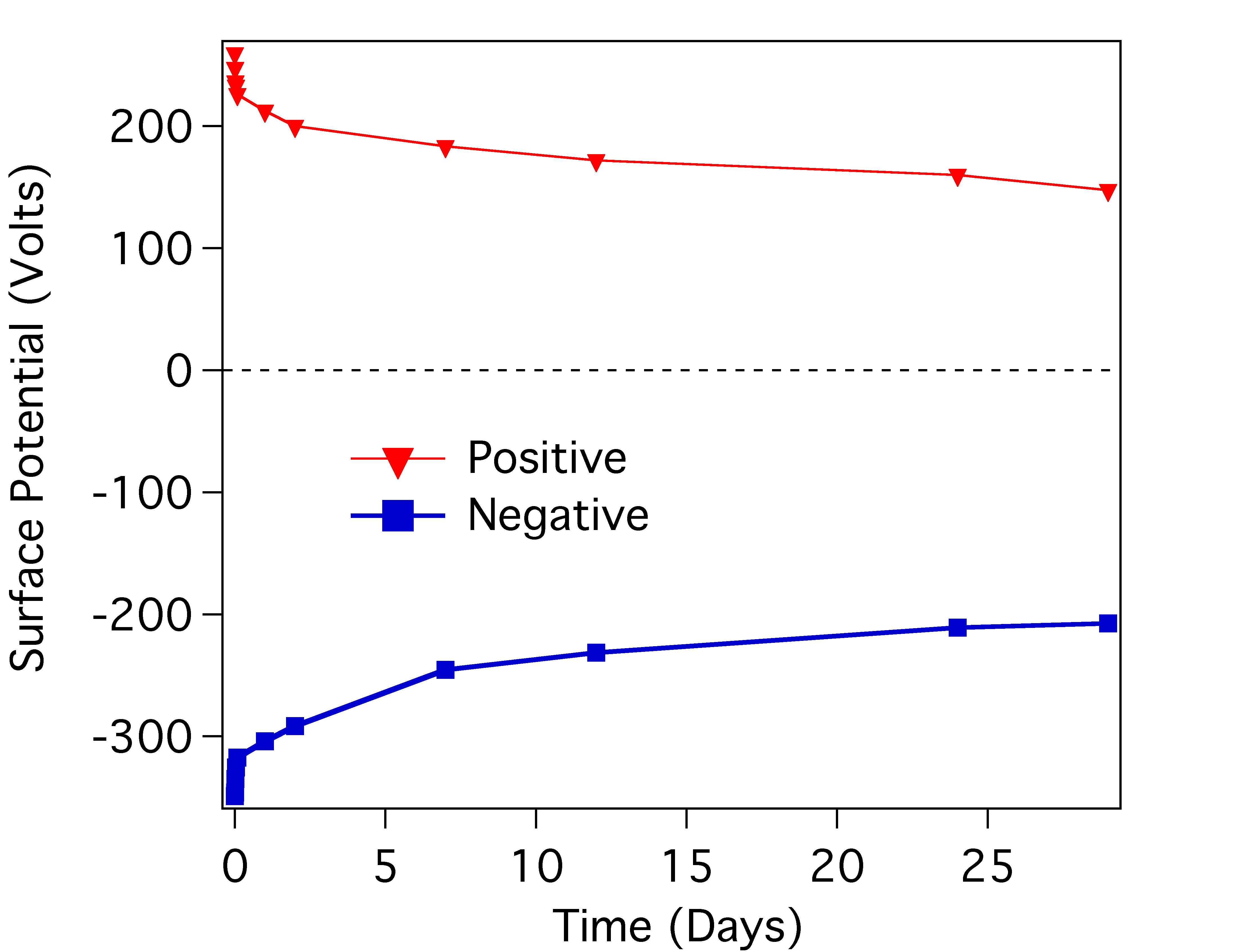}
\caption{Potential stored in Novec after a) charging at constant grid voltage of $\pm$400 V as a function of time for an 5 $\mu$m thick film, b) charging for a fixed time (5 minutes) at incremental grid voltages for film thickness of approximately 0.1, 2, and 5 $\mu$m samples and c) charging at the two high grid voltage of 400 and -400 V and then monitoring the stability of stored charge as a function of time for a 5 $\mu$m thickness film (samples left in air).}
\label{Stability}
\end{figure*}

\begin{figure*}[tb]
\includegraphics[width=0.65\columnwidth]{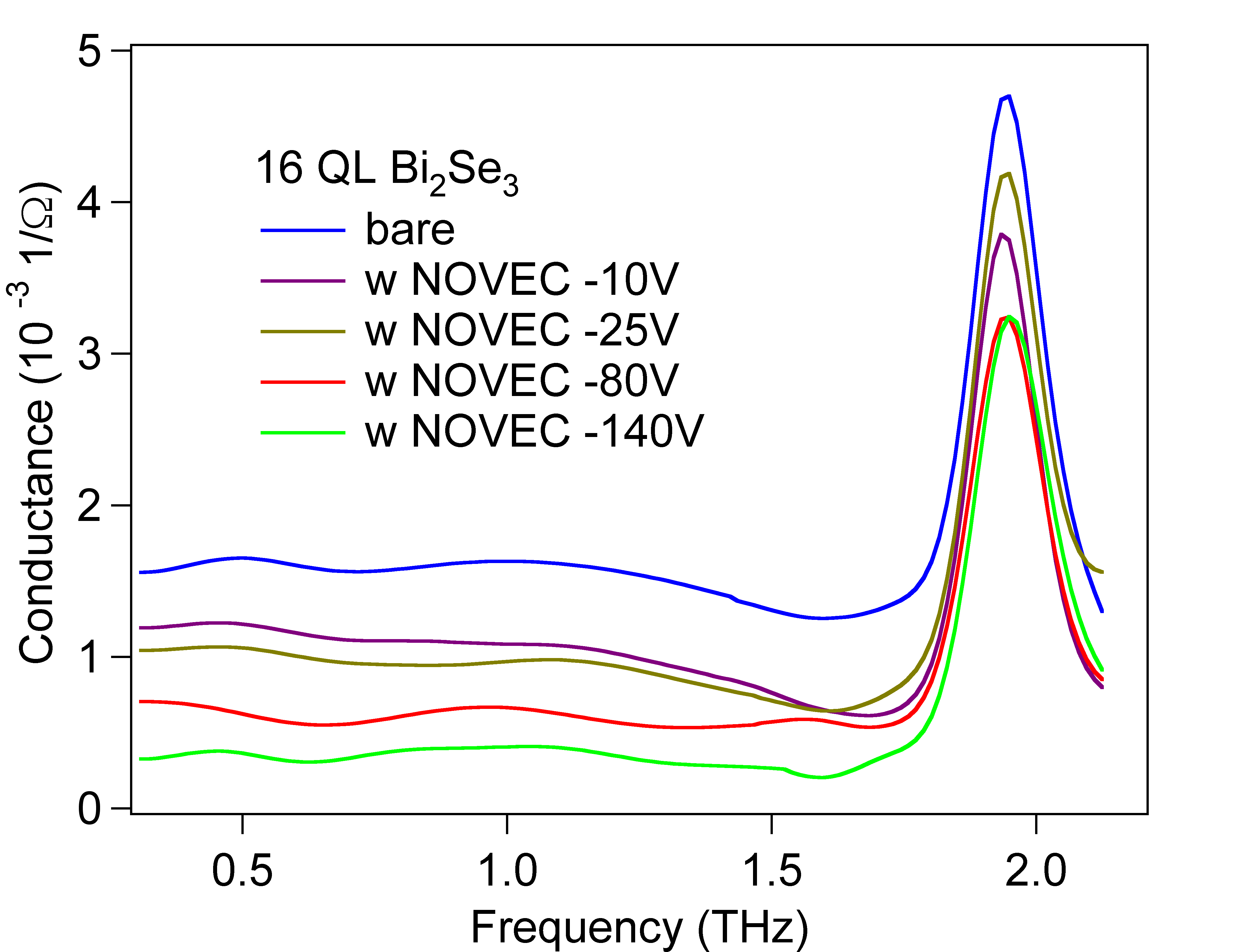}
\includegraphics[width=0.65\columnwidth]{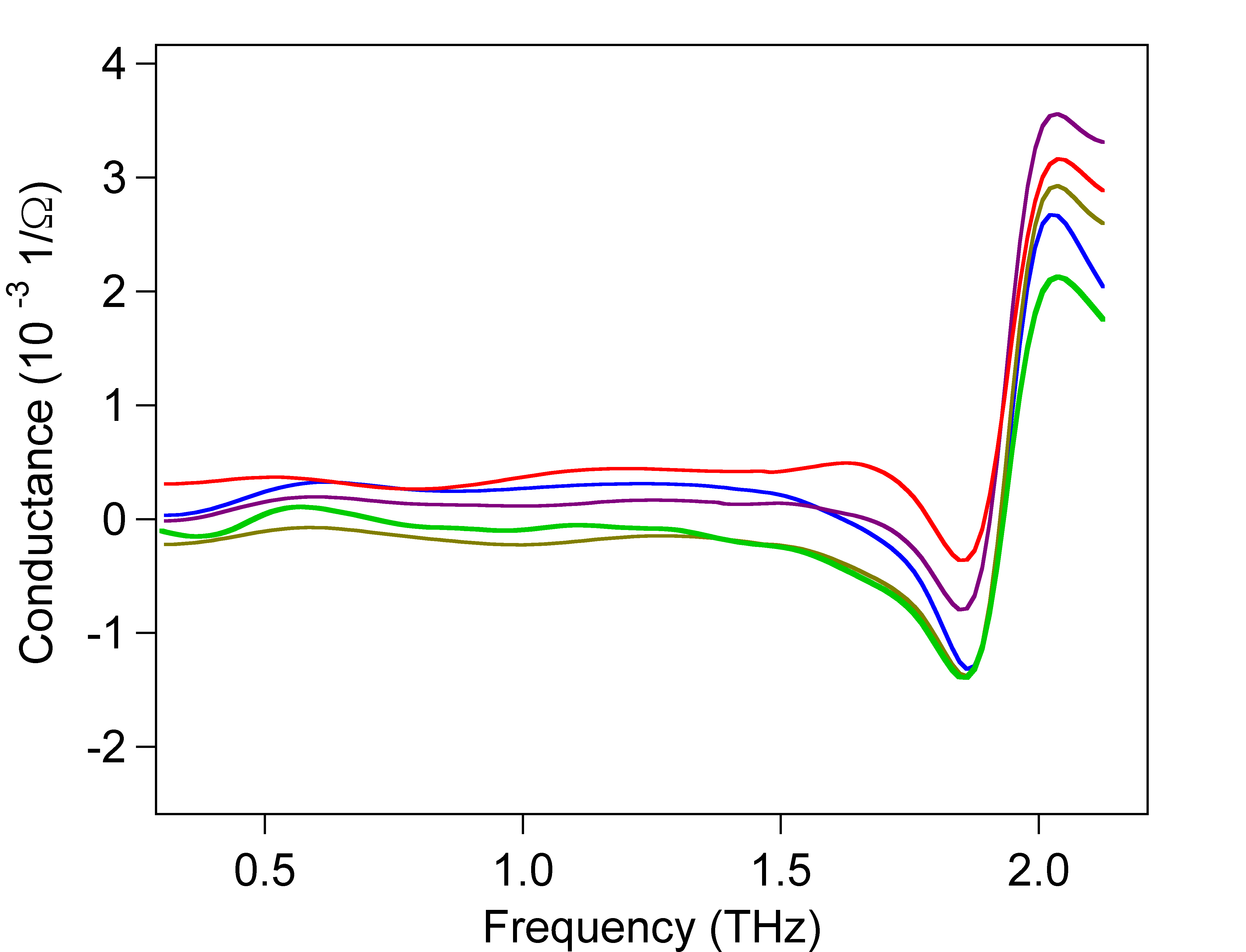}
\includegraphics[width=0.65\columnwidth]{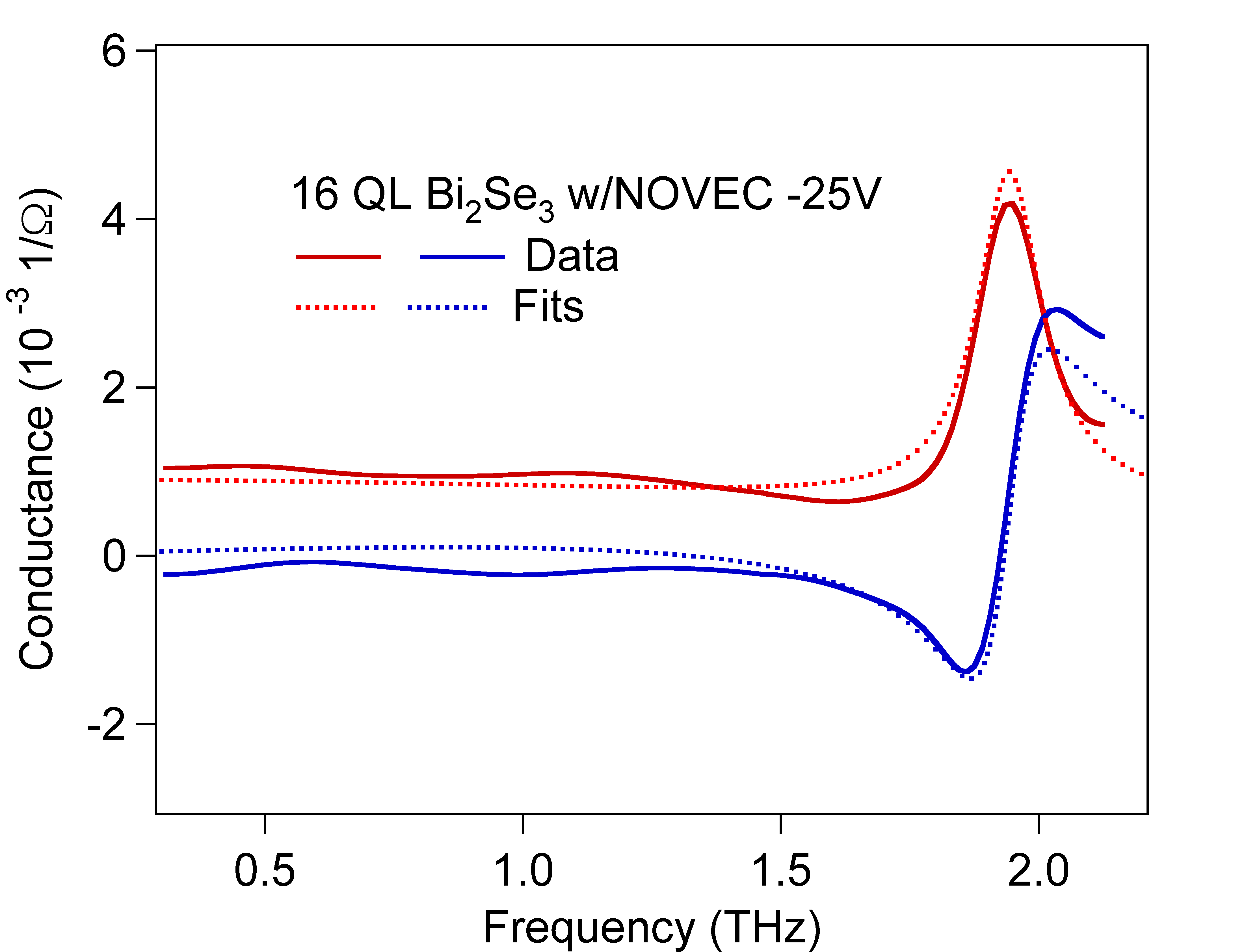}
\caption{ THz data taken at 5 K.  a) Real and b) imaginary parts of the frequency-dependent THz range conductance of Bi$_2$Se$_3$ films with an approximately 0.5 $\mu$m encapsulation layer from multiple spun castings of Novec deposited on top at different levels of charging. c) Real and imaginary parts compared directly for -25 V charging.}
\label{ ThzSpectra}
\end{figure*}

Charging was carried out in an ambient atmosphere for specific time periods and different grid voltages to determine the time evolution and maximum deposited static voltage for each multilayer.    In Fig. \ref{SurfaceCharge}a we show the surface charge voltage measured for polystyrene (PS), polystyrene with 3$\%$ chloromethylstyrene comonomer (PS-Cl), and polystyrene with 3$\%$ C60-methyl side chains (PS-C60) spin-coated onto a gold film.  One can see that grounding increased the surface voltage that could be established.  The stored surface potential saturates at a particular value for some grid voltage (in the tens of volts for 0.5 $\mu$m thick spin-cast polymers).  The time scales are consistent with a measured RC time constant of 20-150 seconds for the polymer films.  Conclusions drawn from this are that both of the functionalized side chains contributed to the charge storage ability, about half of the maximum voltage that could be stored across the entire sample could be localized between the top and bottom polymer surfaces.  About one third more voltage could be maintained if the sample were heated to 80$^\circ$ C during charging.   As shown in  Fig. \ref{SurfaceCharge}b Novec had even better properties in holding charge than PS-Cl.   There one can see that with -1500 V applied grid voltage as comparing PS-Cl Novec deposited on Bi$_2$Se$_3$ clearly showed superior abilities to charge to larger voltages and in shorter times.   We speculate that as Novec is fluorinated and otherwise lacking in polar functional groups, any sites in which free electrons or holes can be trapped are highly stabilized against being quenched by environmental water molecules and also have such high-barrier migration pathways that they are not easily discharged.

Given its superior properties, we proceeded to do an in-depth study of the capacity of Novec to store charge and exhibit a stable surface potential on TIs.  Fig. \ref{Stability} shows various tests for the long term stability of the stored potential in Novec.   Fig. \ref{Stability}a shows the surface potential after charging at constant grid voltage as a function of time. Fig. \ref{Stability}b shows the resulting potential stored in Novec after charging for a fixed time (5 minutes) at incremental grid voltages. Fig. \ref{Stability}c shows the resulting potential stored in Novec after charging at single high grid voltage and then monitoring the stability of stored charge as a function of time (samples left in air) over the course of almost one month.  One can see that Novec exhibits excellent properties in terms of large voltages that can be stored over long time scales.

\section{Effectiveness in modifying charge conduction}

 The complex conductivity was determined by time-domain THz spectroscopy (TDTS).   A femtosecond laser pulse is split along two paths and excites a pair of photoconductive `Auston'-switch antennas.    A broadband THz range pulse is emitted by one antenna, transmitted through the TI film, and measured at the other antenna.  By varying the length-difference between the two paths, one uses the split off laser beam to optically gate the second switch and map out the entire electric field of the transmitted pulse as a function of time.  Taking the ratio of the Fourier transform of the transmission through the TI on a substrate to that of a bare substrate gives the  complex transmission function.  We then invert the transmission to obtain the complex conductivity via the standard formula for thin films on a substrate:  $\tilde{T}(\omega)=\frac{1+n}{1+n+Z_0\tilde{\sigma}(\omega)d} $ where $n$ is the index of refraction of the sapphire substrate.  By measuring both the magnitude and phase of the transmission, this inversion to conductivity is done directly and does not require Kramers-Kronig transformation.    As a characterization tool TDTS has the advantages of being non-destructive and non-contact.   Moreover it accesses the conventionally difficult to utilize THz part of the electromagnetic spectrum.

We demonstrate the effectiveness of a monolithic single-component charged electret in extracting charge carriers from bismuth selenide utilizing THz measurements\cite{Wu16a}.  In the THz spectra in Fig. \ref{ ThzSpectra}, there is a Drude component peaked at zero frequency and a phonon peak at $\sim$ 1.9 THz as observed extensively previously \cite{ValdesAguilarPRL12,Wu16a,WuNatPhys13}. These features manifest themselves in both the real and imaginary parts of the conductivity.  As done previously, we fit these spectra to a model where we fit the spectra to a Drude peak, a Drude-Lorentz peak for the phonon and a $\varepsilon_\infty$ term that accounts for the polarizability of the lattice.  The area of the peak assigned to the Drude conductance is proportional to carrier density over effective mass. Therefore Drude peaks with lower area means lower carrier density.  We fit both the real and imaginary parts of the conductance simultaneously to a Drude transport model.   Using the known band structure of Bi$_2$Se$_3$ \cite{Wu16a, WuPRL2015}, we can extract out the carrier density and also the chemical potential.  It is reasonable to assume that the charge polymer layer does not modify the essential aspects of the band structure as it is spin coated on and not bonded in any fashion.

\begin{figure}[tb]
\includegraphics[width=0.95\columnwidth,angle=0]{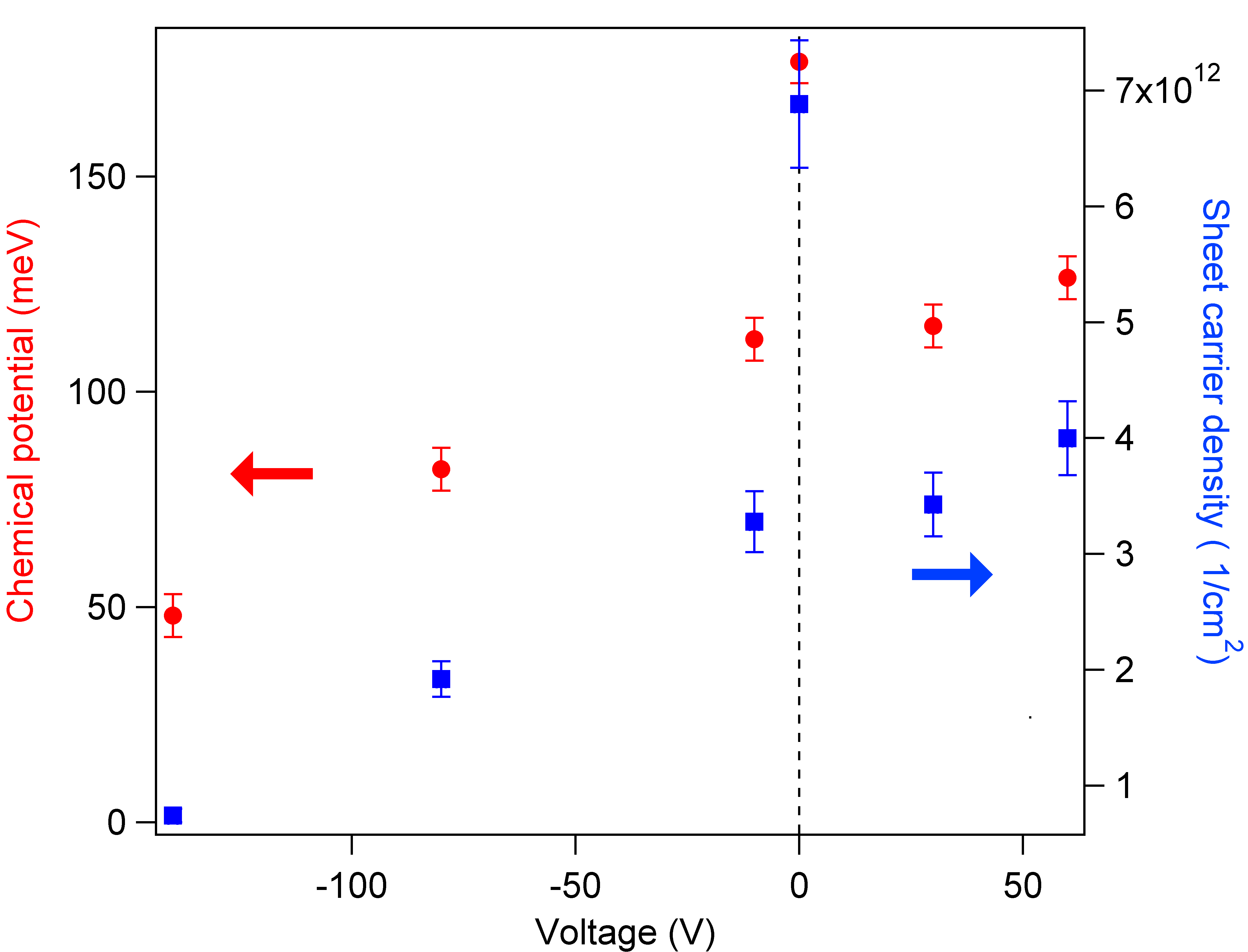}
\caption{Chemical potential and charge density of Novec spin-cast films vs. charging voltage as measured from THz experiments at 5K.   The high points at $V=0$ represent ``bare" films.}
\label{Density}
\end{figure}

We found that Bi$_2$Se$_3$  with a negatively charged polymer layer has the smallest conductance area and the positively charged one barely changes the conductance, indicating that negatively charged polymers deplete the carrier density of surface states of  Bi$_2$Se$_3$ while the positive one slightly increases the carrier density.  This polarity is consistent with charge stored as a dipole, with one pole in the dielectric and the other pole in the  Bi$_2$Se$_3$ or at its polymer interface, where electrons would be trapped (Fig \ref{ChargingCartoon}). In Fig. \ref{Density} we can see that immediately upon spin coating, the charge density of the Bi$_2$Se$_3$ film falls and we found that Novec usually automatically traps $\sim$ -10 V.  It is possible that Novec introduces an interface dipole that itself depletes some charges, or counteracts some pre-existing surface dipole or chemical species that stabilizes a certain quantity of charge in the uncoated TI sample.

Charging was done under grounding conditions.  For large positive applied charges, the charge density increases slightly, but for negative charges the density decreases dramatically.   Using the known band structure, the chemical potential  of the film as a function of the charging potential can be derived \cite{WuPRL2015, Wu16a}.   For the largest charging potentials, the chemical potential can be tuned to within 50 meV of the Dirac point.  By using the band dispersion and measuring the Drude transport life time in THz conductivity, we can estimate the mobility at 5 K is $\sim$ 1600 cm$^2$/Vs at 5K.  We repeated charging experiments on three spin-coated Novec-Bi$_2$Se$_3$ samples and observed voltage dependence as the one shown in Fig. 5.  We also measured two Bi$_2$Se$_3$ samples with drop-cast Novec a few micrometer thick and saw similar behavior.   For such samples we observed similar depletion, but because the analysis is more complicated due to the optically thicker polymer layer, we show here only the spin-coated results.

\section{Conclusions }

We have demonstrated the ability to reduce the carrier concentration of thin films of the topological insulator (TI) Bi$_2$Se$_3$ by utilizing a novel approach, namely non-volatile electrostatic gating via corona charging of electret polymers.  We demonstrate that a perfluorinated electret polymer is a highly effective static gate material to modulate topological insulator charge density.    Sufficient electric field can be retained by this polymer to accomplish significant electron depletion of  Bi$_2$Se$_3$, shifting the Fermi level towards the Dirac point and lowering the surface chemical potential into the bulk band gap.  The  Bi$_2$Se$_3$ is stabilized in the intrinsic regime while its electron mobility is enhanced.  Our work represents the first use of a charged polymer gate for modulating TI charge density.

\section{Acknowledgements }

 The topological insulator preparation and physical study was supported by NSF DMR-1005398 with additional support to SO by a Gordon and Betty Moore Foundation, EPiQS Initiative Grant GBMF4418.  The charged polymer selection, process development, and characterization were supported by the Department of Energy, Office of Basic Energy Sciences, Grant Number SPACE DE-FG02-07ER46465..  RMI and LW contributed equally to this work.

\bibliography{TopoIns}

\end{document}